\documentclass[showpacs,showkeys,preprintnumbers]{revtex4}%
\usepackage{amsfonts}
\usepackage{amsmath}
\usepackage{graphicx}
\usepackage{float}
\usepackage{dcolumn}
\usepackage{bm}
\usepackage{amssymb}%
\begin{document}
\title{Decay of the pseudoscalar glueball into vector, axial-vector, scalar and pseudoscalar mesons}
\author{Walaa I.\ Eshraim}
\affiliation{New York University Abu Dhabi, Saadiyat Island, P.O. Box 129188, Abu Dhabi, U.A.E}

\begin{abstract}
We resume the investigation of the ground-state pseudoscalar glueball, $J^{PC}=0^{-+}$, by computing its two- and three-body decays into vector and axial-vector quark-antiquark meson fields additional to scalar and pseudoscalar mesons through the construction of an interaction chiral Lagrangian that produces these decays. We evaluate the branching ratio, via a parameter-free calculation, by setting the mass of the pseudoscalar glueball to $2.6$ GeV as predicted by lattice QCD simulations. We duplicate the computation for the branching ratios for a pseudoscalar glueball mass $2.37$ GeV which matches to the measured mass of the resonance $X(2370)$ in the BESIII experiment. The present channels and states are potentially reached and are interesting for the running BESIII and Belle-II experiments and the planned PANDA experiment at  FAIR/GSI which will be able to detect the pseudoscalar glueball within the accessible energy range.  
\end{abstract}

\pacs{12.39.Fe, 13.20.Jf, 12.39.Mk, 12.38.-t}
\keywords{chiral Lagrangians, (axial-)vector, (pseudo) scalar mesons, pseudoscalar glueball, BES, PANDA}\maketitle

\section{Introduction}

Glueballs, the composite particles containing gluons without valence quarks, are predicted by the confinement properties of Quantum chromodynamics (QCD) \cite{bag-glueball} and the non-Abelian nature \cite{Nakano} of the $SU(3)_c$ colour symmetry by virtue of the gauge fields of QCD, the gluons,  self-interaction and strong vacuum fluctuations. A scalar glueball, $J^{PC}=0^{++}$, is the ground state and its mass range is estimated to be from $1000$ to $1800$ MeV. That followed by a pseudoscalar glueball, $J^{PC}=0^{-+}$, at higher mass. Up to now, glueballs remain experimentally undiscovered \cite{Jia} because of their mixing with ordinary meson states, and no meson is listed by the Particle Data Group (PDG) \cite{PDG} to be unambiguous of predominant glueball nature. Therefore, the search for glueballs witnessed extensive and intensive investigations by both theoretical and experimentally studies \cite{Morningstar, Gregory, EshraimG, EshraimSGN, staniG, EshraimTH, EshraimGS, EshraimEG, Sun, Hub}.  Actually, ten scalars including the isoscalars $f_0(1370)$, $f_0(1500)$, and $f_0(1710)$ are observed. The last resonance is classified to be predominantly a scalar glueball, according to the result of the numerical approach of the lattice QCD \cite{Morningstar2, Lattice} and effective approaches \cite{staniG, scalars, Lee}. Theoretical studies of glueballs are performed by nonperturbative approaches, the flux tube model \cite{Isgur}, constituent models \cite{Boul}, the holographic approach \cite{Rab}, effective chiral models \cite{EshraimG, staniG, EshraimGS,  EshraimEG} and lattice QCD simulations \cite{Morningstar2, Lattice} which reach the spectrum of glueball states below $5$ GeV, meaning that these simulations play an important role in the investigation of the low energy strong interaction phenomena. In the quenched approximation, lattice QCD simulation computed the masses of glueballs  \cite{Morningstar, Morningstar2}. For example, it predicted the mass of the ground-state pseudoscalar glueball around $2.6$ GeV and the mass of the first excited state of the pseudoscalar glueball around $3.7$ GeV. On the other hand, the production rates of glueballs in the $J/\psi$ radiative decays \cite{Yang1, Yang2} represent an additional role in determining the glueballs owing to the gluon-rich environment.  In the quenched approximation, the pure gauge glueballs are well-defined hadron states. Therefore, the electromagnetic form factors of $J/\psi$ radiatively decaying into glueballs are directly extracted from the calculation of the matrix elements of the electromagnetic current between glueballs and $J/\psi$. The BESIII collaboration studied the process $J/\psi \rightarrow \gamma \eta' \pi^+\pi^-$ and observed in the $ \eta' \pi^+\pi^-$ decay the resonance $X(2370)$ with quantum number $J^{PC}=0^{-+}$ \cite{bes}. This resonance has the same quantum number of a pseudoscalar glueball and lies in the mass range of the pseudoscalar glueball in lattice QCD prediction. That leads us to suggest the assignment of the pseudoscalar glueball also to the $X(2370)$ in our work previously published in Ref. \cite{EshraimG} and references therein. There are also several candidates for the pseudoscalar glueball as seen for example in Ref. \cite{Nekrasov}, where the lower pseudoscalar glueball state is suggested to be the upper iota $\eta(1490)$. Moreover, the resonance $I(1440)$ is required to be a pseudoscalar glueball in Ref. \cite{Genz} out of the $\eta-\eta'-I(1440)-\eta_C$ mixing investigation. The phenomenology of the pseudoscalar glueball is studied in a family of finite-width Gaussian sum rules upon a correction from instanton–gluon interference to the correlation function in Ref. \cite{wang}, where the authors concluded that the interference contribution is gauge-invariant. 

The present study of the decay properties of the ground state pseudoscalar glueball (denoted as $\tilde{G}$) is based on the chiral symmetric model of low-energy QCD called the extended Linear Sigma Model (eLSM) \cite{dick}. It contains all quark-antiquark mesons with (pseudo)scalar and (axial)vector as well as a scalar and a pseudoscalar glueball and implements the symmetries of the QCD and their breaking. The eLSM is interesting for the study the hadron phenomenology. One can see in Ref. \cite{Law-Energy} that the eLSM compatible with chiral perturbation theory for what concerns low-energy pions (most notably, pion-pion scattering). In detail, Ref. \cite{dick} achieved a good description of PDG data by a fit to various experimental quantities. That fit allowed to fix the parameters of the model, subsequently making other predictions/postdiction. Both conventional $\overline{q}q$-states and various non-conventional gluonic mesons were already studied in the eLSM.  The phenomenology of the light mesons \cite{dick} and excited mesons \cite{excitedmesons} have been nicely described, as well as the properties of the open and hidden charmed mesons \cite{EshraimC, Eshraimeta}. As a consequence of dilatation invariance and its anomalous, the scalar glueball appears naturally in the eLSM which allowed to study the vacuum properties of the scalar glueball \cite{stani, staniG}. The eLSM has been also applied to the hadronic decays of the pseudoscalar glueball(s) \cite{ EshraimG, EshraimGS, EshraimEG} and the vector glueball \cite{F.julia}. As an additional advantage, in Ref.\cite{gallas, Olb, Olb2, lakas}, the eLSM has been successfully applied in the baryonic sector within the so-called mirror assignment which the predictions turn out to be in agreement with the experimental data for pion-nucleon scattering and baryonic decays. The eLSM was also employed at a finite temperature \cite{Kovacs, Tawfik} and density \cite{lakas, Heinz}, to describe the chiral phase transition in the medium. On the other hand, the inclusion of hybrid mesons was presented in Ref. \cite{hybrid}.

 Within the eLSM \cite{EshraimG}, we have analysed the decay of the ground-state pseudoscalar glueball into scalar and pseudoscalar mesons and found that the channel $KK\pi$ is dominant and the $KK\eta$ and $KK\eta'$ decay modes are sizable. Moreover, the hadronic decays of the pseudoscalar glueball into nucleons were computed \cite{EshraimSGN} and into (pseudo)scalar mesons and their first excited state Ref. \cite{EshraimEG}. Furthermore, the decay properties of the first excited pseudoscalar glueball have been also studied in Refs. \cite{EshraimGS, EshraimEG}. These efforts on the pseudoscalar glueball and its first excitation properties are important in the comprehension of the non-perturbative behaviour of QCD and useful in searching for the pseudoscalar glueball in future experiments. Based on that, we are interested in continuing our investigations on the properties of glueballs.

In this paper,  we use both masses $M_{\tilde{G}}=2.37$ GeV relevant to the BESIII experiment candidate and $M_{\tilde{G}}=2.6$ GeV predicted by the lattice QCD simulation,  to calculate the decay widths of the pseudoscalar glueball in the framework of the constructed effective model so to connect the pseudoscalar glueball, $gg$, to $\overline{q}q$ vector and axial-vector mesons in addition to scalar and pseudoscalar mesons. This work is a further step in our investigations of the pseudoscalar glueball and predicts new decay channels including vector and axial-vector mesons. We can thus compute the widths for the decays $\tilde{G}\rightarrow PS$, $\tilde{G}\rightarrow PV$, $\tilde{G}\rightarrow PPP$,  $\tilde{G}\rightarrow PPA$, $\tilde{G}\rightarrow PPV$, and $\tilde{G}\rightarrow PSV$, where $P,\,S,\, V$, and $A$ stand for pseudoscalar, scalar, vector, and axial-vector quark-antiquark states, respectively. The pseudoscalar field $P$ corresponds to the well-known light mesons $\{\pi, K, \eta, \eta'\}$ and the scalar $S$ corresponds to the scalars above $1$ GeV: $\{a_0(1450),\,K^*_0(1430),\,f_0(1370),\,f_0(1500)\}$, while the vector state $V$ refers to $\{\rho(770),\, K^*(892),\, \omega,\,\phi\}$ and the axial-vector $A$ refers to $\{a_1(1260), f_0(1285),\, f_1(1420),\,K_1(1200)\}$. The results are presented as branching ratios in order to disregard the unknown coupling constant. The present results confirm all channels which were already predicted earlier in Ref. \cite{EshraimG}: the decays of the pseudoscalar glueball into scalar and pseudoscalar mesons. This is particularly interesting, as it may help the community understand the hadron spectrum and the search for glueballs in future experiments. The present investigation for the two- and three-body decays of the ground-state pseudoscalar is a useful guideline for both the running  BESIII/(Beijing, China) and Belle-II/(Tsukuba, Japan) experiments and the planned PANDA experiment at the FAIR/(GSI, Germany)  \cite{panda} which are able to measure the proposed channels.

The present paper is organized as follows. In Sec. II we present the chiral multiplets. Then, in Sec. III we introduce the constructed effective  Lagrangian which describes the two- and three-body decays of the ground-state pseudoscalar glueball into vector, axial-vector, scalar and pseudoscalar quark-antiquark degrees of freedom, allowing for the branching ratios prediction for the decays. Finally, in Sec. IV we present the conclusions.

\section{Chiral Multiplets}

 In this section, we present the quark-antiquark fields which represent the pseudoscalar glueball decay products. On the basic ingredients of the eLSM, the (pseudo)scalar and (axial-)vector field mesons are presented below, where the chiral combinations were properly taken into the account.\\

The multiplet of the scalar $S^a$, and the pseudoscalar $P^a$, quark-antiquark states is introduced \cite{dick} as
\begin{equation}
\Phi=(S^{a}+iP^{a})t^{a}=\frac{1}{\sqrt{2}}\left(
\begin{array}
[c]{ccc}%
\frac{(\sigma_{N}+a_{0}^{0})+i(\eta_{N}+\pi^{0})}{\sqrt{2}} & a_{0}^{+}%
+i\pi^{+} & K_{0}^{*+}+iK^{+}\\
a_{0}^{-}+i\pi^{-} & \frac{(\sigma_{N}-a_{0}^{0})+i(\eta_{N}-\pi^{0})}%
{\sqrt{2}} & K_{0}^{*0}+iK^{0}\\
K_{0}^{*-}+iK^{-} & \bar{K}_{0}^{*0}+i\bar{K}^{0} & \sigma_{S}+i\eta_{S}%
\end{array}
\right)  \; , \label{phimatex}%
\end{equation}

where $t^a$ are the generators of the group $U(N_f)$. The multiplet $\Phi$ transforms under $U_{L}(3)\times U_{R}(3)$ chiral transformations as $\Phi\rightarrow U_{L}\Phi U_{R}^{\dagger}$, whereas $U_{L(R)}=e^{-i\Theta^a_L(R)^{t^a}}$ are $U(3)_{L(R)}$ matrices, and under the charge conjugation $C$ as $\Phi\rightarrow \Phi^T$ as well as under the parity $P$ as $\Phi(t,\overrightarrow{x})\rightarrow \Phi^\dag(t,\overrightarrow{x})$.\\ 

The vector $V^{a}$ and axial-vector $A^{a}$, degree of freedom, are presented \cite{dick} in the following left- and right-handed matrices, $L_\mu$ and $R_\mu$, as

\begin{equation}\label{4}
L_\mu=(V^a+i\,A^a)_{\mu}\,t^a=\frac{1}{\sqrt{2}}
\left(%
\begin{array}{ccc}
  \frac{\omega_N+\rho^{0}}{\sqrt{2}}+ \frac{f_{1N}+a_1^{0}}{\sqrt{2}} & \rho^{+}+a^{+}_1 & K^{*+}+K^{+}_1  \\
  \rho^{-}+ a^{-}_1 &  \frac{\omega_N-\rho^{0}}{\sqrt{2}}+ \frac{f_{1N}-a_1^{0}}{\sqrt{2}} & K^{*0}+K^{0}_1 \\
  K^{*-}+K^{-}_1 & \overline{K}^{*0}+\overline{K}^{0}_1 & \omega_{S}+f_{1S} \\
\end{array}%
\right)_\mu\,,
\end{equation}
and
\begin{equation}\label{5}
R_\mu=(V^a-i\,A^a)_\mu\,t^a=\frac{1}{\sqrt{2}}
\left(%
\begin{array}{ccc}
  \frac{\omega_N+\rho^{0}}{\sqrt{2}}- \frac{f_{1N}+a_1^{0}}{\sqrt{2}} & \rho^{+}-a^{+}_1 & K^{*+}-K^{+}_1 \\
  \rho^{-}- a^{-}_1 &  \frac{\omega_N-\rho^{0}}{\sqrt{2}}-\frac{f_{1N}-a_1^{0}}{\sqrt{2}} & K^{*0}-K^{0}_1  \\
  K^{*-}-K^{-}_1 & \overline{K}^{*0}-\overline{K}^{0}_1 & \omega_{S}-f_{1S} \\
\end{array}%
\right)_\mu\,.
\end{equation}
Under $U_{L}(3)\times U_{R}(3)$ chiral transformations, the multiplets  $L_\mu$ and $R_\mu$ transform as $L_{\mu}\rightarrow U_{L} L_{\mu}U_{L}^{\dag}$ and $R_{\mu}\rightarrow U_{R}L_{\mu}U_{R}^{\dag}$, respectively. In the present investigation, we are interested in studying the hadronic decays of the pseudoscalar glueball field $\tilde{G}$ which is chirally invariant and transforms under charge conjugation as $\tilde{G}\rightarrow \tilde{G}$ and under the parity $P$ as $\tilde{G}(t,\overrightarrow{x})\rightarrow - \tilde{G}(t,\overrightarrow{x})$. Consequently, these transformation properties of the multiplets $\Phi,\, L_\mu,\, R_\mu$ and the pseudoscalar glueball $\tilde{G}$ have
been used to construct the below effective invariant Lagrangian (\ref{lag1}) and the extended Linear Sigma Model (eLSM), see Appendix A and Ref. \cite{dick} as well.

 The identification of the quark-antiquark fields in the present model (\ref{lag1}) with the physical resonances presented in details in Ref. \cite{dick}, is straightforward in the light (pseudo)scalar and (axial-)vector states with mass $\lesssim 2$ GeV. The Pseudoscalar sector $P^a$ includes the pion isotriplet $\overrightarrow{\pi}$, the kaon isodoublet $K$ \cite{PDG}, and the isoscalar fields $\eta_{N}\equiv\left\vert \bar{u}u+\bar{d}d\right\rangle /\sqrt{2}$ and $\eta_{S}\equiv\left\vert \bar{s}s\right\rangle $, which represent the non-strange and strangeness mixing
components of the physical states $\eta$ and $\eta^{\prime}$ \cite{PDG} with mixing angle  $\varphi\simeq-44.6^{\circ}$ \cite{dick}.%
\begin{equation}
\eta=\eta_{N}\cos\varphi+\eta_{S}\sin\varphi,\text{ }\eta^{\prime}=-\eta
_{N}\sin\varphi+\eta_{S}\cos\varphi\,. \label{mixetas}%
\end{equation}
The scalar sector $S^a$ contains the isotriplet  $\vec{a}_{0}$ which refers the
physical resonance $a_{0}(1450)$ and the kaon field $K_0^*$ which is assigned
to the physical isodoublet state $K_{0}^{\star}(1430).$ In the scalar-isoscalar sector, the non-strange bare field $\sigma_{N}$, the bare strange field  $\sigma_{S}$ and the scalar glueball $G$ mix and generate the three physical resonances  $f_{0}(1370)$ , $f_0(1500)$, and $f_{0}(1710)$ through the following mixing matrix as constructed in Ref. \cite{staniG}:
\begin{equation}\label{scalmixmat}
\left(%
\begin{array}{c}
 f_0(1370) \\
 f_0(1500)   \\
 f_0(1710)\\
\end{array}%
\right)=\left(%
\begin{array}{ccccc}
 -0.91 & 0.24 & -0.33\\
 0.30 & 0.94  &-0.17\\
 -0.27 & 0.26 & 0.94\\
\end{array}%
\right)\left(%
\begin{array}{c}
 \sigma_N\\
  \sigma_S\\
  G\\
\end{array}%
\right).
\end{equation}
However, the scalar-isoscalr fields, $f_{0}(1370)$, $f_{0}(1500)$, and $f_{0}(1700)$ are predominantly described by the bare
configuration$\equiv\left\vert \bar{u}u+\bar{d}d\right\rangle /\sqrt{2}$ state, $\bar{s}s$ states and a glueball $gg$ state, respectively. \\
We now turn to the assignment of the (axial-)vector states. The vector sector $V^a$ contains the iso-triplet field $\overrightarrow{\rho}$, the kaon states $\overrightarrow{K}^*$, and the isoscalar states $\omega_N$ and $\omega_S$ which are assigned to the $\rho(770)$, $K^*(892)$, $\omega$ and $\phi$ mesons, respectively \cite{dick}. Note that the mixing between the strange and nonstrange isoscalars vanishes in the extended linear sigma model eLSM \cite{dick}, whereas this mixing is so small as obtained in Ref. \cite{klem}. In the end, for the axial-vector sector $S^a$, the isotriplet $\overrightarrow{a}_1$, the isoscalar fields $f_{1N}$ and $f_{1S}$ correspond to the resonances $a_1(1260)$, $f_1(1285)$ and $f_1(1420)$ respectively. However, the four kaon states $K_1$ refer predominantly to the resonance $K_1(1200)$ and could also refer to $K_1(1400)$ because of the mixing between the pseudovector states and axial-vector states \cite{Hat}.

\section{Decay of the pseudoscalar glueball into conventional mesons}

We consider a chirally invariant Lagrangian which couples the ground-state pseudoscalar glueball $\tilde{G}\equiv\left\vert gg\right\rangle$
with quantum numbers $J^{PC}=0^{-+}$ to the quark-antiquark vector, axial-vector, scalar and pseudoscalar field mesons

\begin{equation}
\mathcal{L}_{eLSM,\,\tilde{G}}^{int}= i\, c \,\tilde{G}\,\mathrm{Tr} \left[L_\mu \left(\partial^\mu \Phi\,.\, \Phi^\dag+\Phi\, .\, \partial^\mu\Phi^\dag \right)-R_\mu \left(\partial^\mu \Phi^\dag\,.\, \Phi + \Phi^\dag\, .\, \partial^\mu\Phi \right)\right]  \text{ ,}
\label{lag1}%
\end{equation}

which is invariant under $U(3)_R\times U(3)_L$, C, and $P$ transformations. The coupling constant $\alpha$ is an unknown coupling constant and has a dimension of Energy$^{-3}$. Based on Ref. \cite{Adrian}, we constructed this model where the heterochiral $\Phi$ and $\Phi^\dag$, involves (pseudoscalar) mesons, coupled to the homochiral $L_\mu$ and $R_\mu$, consisting of (axial-)vector mesons, through only structures which contain derivatives of $\Phi$. This Lagrangian describes the two- and three-body decays of the pseudoscalar glueball. Notice that the two-body decays appear only through the condensate and the interesting thing is that it does not lead to the two-body decays for the nonet of chiral partners $A^a$.

According to the validity of the joint model
\begin{equation}
\mathcal{L}_{eLSM}+\mathcal{L}_{eLSM,\,\tilde{G}}^{int}\,,
\end{equation}
the pseudoscalar glueball in the present work has a mass of about $2.37$ GeV coinciding to a claimed BESIII experiment candidate and of $2.6$ GeV from the lattice QCD prediction, while the eLSM is a low-energy chiral model valid up to $1.7$ GeV. Therefore, it should be accepted that this model is suited to calculate exclusively the decays of the field $\tilde{G}$. For instance in Refs. \cite{EshraimC, Eshraimeta}, the eLSM has found to be applicable to study the phenomenology of the heavy charmed mesons, which concerns the calculation of masses and large-$N_c$ dominant decays although it could be far from the natural domain of chiral symmetry. Moreover, the employed effective models in Ref. \cite{F.julia} have been used in the decay modes of the vector glueball with a mass of about $3.8$ GeV. The present approach is consequently expected to be reliable within a similar accuracy, even if it is proposed here to test the decays of an (as of yet) unidentified glueball. From the Refs. \cite{EshraimGS, EshraimEG} and refs. therein, we can prove the validity of the employed effective chiral model in Eq. (\ref{lag1}) to study the decay modes of the pseudoscalar glueball. The novelty of our approach is that the qualitative outcomes do not depend on the precise value of the input parameters. 

The lagrangian (\ref{lag1}) shows that the pseudoscalar glueball state of such high mass could decay, which is not the only and first case, that appears in this field of investigations. There were widely used models which couple one heavy field to light mesons as seen for instance in Refs. \cite{EshraimGS, EshraimEG, Eshraimeta, F.julia, Close-H, Gutsche-H, Escribano-H} and refs. therein. Also the decays of the heavy scalar and pseudoscalar charmonium states, $\chi_{c0}$ and $\eta_C$, are studied by using the eLSM, that gave results in reasonable agreement with experimental data where available, see the details in Ref. \cite{Eshraimeta}, in additional to the accepted investigation of the decays of the heavy vector charmonium state $J/\psi$ into light pseudoscalar mesons in Ref. \cite{Close-H}. It is axiomatic that this basic assumption of effective hadronic models would be tested through only advanced lattice simulations and/or the future experimental discovering of glueballs. Definitely, when we couple the heavy field such as glueballs to the eLSM, we took into account that glueballs are flavour blind and chirally blind. So chiral symmetry, with its spontaneous breaking, does not affect the determination of the pseudoscalar hadronic decays. While up to now there are no data for a direct comparison, the decay ratios can only be predicted with model-dependent, and by neglecting a mixing influence and symmetry breaking terms. The present outcome branching ratios could be useful in the future search of the ground state pseudoscalar glueball.

For computing the decays of the pseudoscalar glueball, one has to perform the shift the scalar-isoscalar fields by their vacuum expectation values $\phi_N$ and $\phi_S$ to implement the effectiveness of  the spontaneous symmetry breaking which takes place ($m_0^2 < 0$) \cite{dick}

\begin{equation}
\sigma_{N}\rightarrow\sigma_{N}+\phi_{N}\text{ and }\sigma_{S}\rightarrow
\sigma_{S}+\phi_{S}\text{ .} \label{shift1}%
\end{equation}
In matrix form:
\begin{equation}
S\rightarrow\Phi_0+S\,\,\, {\rm with}\, \Phi_0= \frac{1}{\sqrt{2}}
\left(%
\begin{array}{ccccc}
 \frac{\phi_N}{\sqrt{2}} & 0 & 0\\
 0 &  \frac{\phi_N}{\sqrt{2}}  & 0\\
 0 & 0 &  \phi_S\\
\end{array}%
\right).
\end{equation}

In addition, we shift also the axial-vector fields to redefine the wave-function renormalization constants of the pseudoscalar fields
\begin{equation}
\vec{\pi}\rightarrow Z_{\pi}\vec{\pi}\text{ , }K^{i}\rightarrow Z_{K}%
K^{i}\text{, }\eta_{j}\rightarrow Z_{\eta_{j}}\eta_{j}\;, \label{psz}%
\end{equation}
whereas $i=1,2,3,4$ indicates the four kaonic fields and $j$ refers to $N$ and $S$. The
numerical values of all the parameters and the renormalization constants of the corresponding wave-functions appearing in the present paper expressions have been fixed in Ref. \cite{dick}. Their values are summarized in Table I.

\begin{table}[H]
\centering
\begin{tabular}
[c]{|c|c|c|c|c|} \hline parameter & value  & renormalization factor & value\\
\hline $\omega_{a_1}= \omega_{f_{1N}}$  & $0.00068$   & $Z_\pi=Z_{\eta_N}$ & 1.70927\\
\hline $\omega_{K^*}$ & $-0.00005\text{ i}$ &  $Z_K$  & 1.60406  \\
\hline  $\omega_{f_{1S}}$ & $0.00056$ & $Z_{\eta_S}$ & 1.53854\\
\hline $\omega_{K_1}$ & $0.00061$ & $Z_{K^*_{0}}$ & 1.00105\\
\hline
\end{tabular}
\caption{Parameters and wave-function renormalization constants.}
 \label{Tab:ren}
\end{table}
The equivalence between the wave-function renormalization constants $Z_\pi$ and $Z_{\eta_N}$ comes out from the  isospin symmetry, accordingly for $\omega_{a_1}$ and $\omega_{f_{1N}}$. 
The corresponding chiral condensates $\phi_N$ and $\phi_S$ read

\begin{align}
\phi_{N}=&Z_{\pi}f_{\pi}=0.158\text{ GeV, }\,\,\,\,\,\,\,\,\,\phi_{S}=\frac{2Z_{K}f_{K}-\phi_{N}%
}{\sqrt{2}}=0.138\text{ GeV}\;,
\end{align}
where the decay constant of the pion is $f_{\pi}=0.0922$ GeV and the kaon is $f_{K}=0.110$ GeV \cite{PDG}. 

After performing the shift operations (\ref{shift1}) and (\ref{psz}) in the Lagrangian\ (\ref{lag1}),  we obtain the relevant tree-level vertices for the decay processes of pseudoscalar glueball $\tilde{G}$.\\

The branching ratios of the pseudoscalar glueball $\tilde{G}$ for the decays into two and three-body, $PS$, $PV$, $PPP$, $PPV$, $PPA$ and $PSV$ are reported in the following tables for two possible choices of the pseudoscalar glueball masses. The choice of the value $2.6$ GeV is a consequence of the central value of a given lattice calculation and $2.37$ GeV according to the obtained candidate by the BESIII experiment. The results are presented relative to the decay width of the pseudoscalar glueball into $\pi\pi \eta$, $\Gamma_{\tilde{G}\rightarrow\pi\pi \eta}$, to eliminate the unknown coupling constant.

\begin{center}%
\begin{table}[H] \centering
\begin{tabular}
[c]{|c|c|c|}\hline
Quantity & Case (i): $M_{\tilde{G}}=2.6$ GeV & Case (ii): $M_{\tilde{G}}=2.37$
GeV\\\hline
$\Gamma_{\tilde{G}\rightarrow K K^*}/\Gamma_{\tilde{G}\rightarrow\pi\pi \eta}$ &
$0.00026$ & $0.00031$\\\hline
$\Gamma_{\tilde{G}\rightarrow a_0\pi}/\Gamma_{\tilde{G}\rightarrow\pi\pi \eta}$ & $0.1913$
& $0.1858$\\\hline
$\Gamma_{\tilde{G}\rightarrow KK_S}/\Gamma_{\tilde{G}\rightarrow\pi\pi \eta}$ & $0.1745$ &
$0.1595$\\\hline
$\Gamma_{\tilde{G}\rightarrow f_0(1370) \eta}/\Gamma_{\tilde{G}\rightarrow\pi\pi \eta}$
& $0.0374$ & $0.0349$\\\hline
$\Gamma_{\tilde{G}\rightarrow f_0(1500) \eta}/\Gamma_{\tilde{G}\rightarrow\pi\pi \eta}$ & $0.00399$ & $0.00325$\\\hline
$\Gamma_{\tilde{G}\rightarrow f_0(1700)\eta }/\Gamma_{\tilde{G}\rightarrow\pi\pi \eta}$ & $0.00265$ &
$0.00134$\\\hline
$\Gamma_{\tilde{G}\rightarrow f_0(1370) \eta^{\prime}}/\Gamma_{\tilde{G}\rightarrow\pi\pi \eta}$& $0.00837$ & $0.00343$\\\hline
$\Gamma_{\tilde{G}\rightarrow f_0(1500)\eta^{\prime} }/\Gamma_{\tilde{G}\rightarrow\pi\pi \eta}$ & $0.00999$ &
$0$\\\hline
\end{tabular}%
\caption{Branching ratios for the decay of the pseudoscalar glueball $\tilde
{G}$ into $PV$ and $PS$.}%
\end{table}%
\end{center}
In Table II, we predict the same channels of the decay of the pseudoscalar glueball into $PS$, which were presented earlier in Ref. \cite{EshraimG} additional to the only new channel $\tilde{G}\rightarrow K K^*$ which describe the validity of the pseudoscalar glueball to decay into the vector meson $K^*$  and the pseudoscalar meson $K$. That confirm all the previous two-body $PS$ channels and the decay of the pseudoscalar glueball into scalar-isoscalar states $f_{0}(1370)$, $f_{0}(1500)$ and $f_0(1700)$ including the full mixing pattern above $1$ GeV.  (For details of the two-body decay calculation, see Appendix B). The two-body decay channels $a_0\pi$ and $KK^*$ are sizable.
Moreover, the two-body decay channel $\tilde{G}\rightarrow KK_S$ can proceed through a sequential instance, $K^*_0(1430)\rightarrow K\pi$, leading to the three-body decay $\tilde{G}\rightarrow KK\pi$. In order to obtain the total three-body decay width for $\tilde{G}\rightarrow KK\pi$, the two- and three-body decay amplitudes of this channel have to be added coherently before taking the modulus square as seen in Ref.\cite{EshraimSGN}.\\

\begin{center}%
\begin{table}[H] \centering
\begin{tabular}
[c]{|c|c|c|}\hline
Quantity & Case (i): $M_{\tilde{G}}=2.6$ GeV & Case (ii): $M_{\tilde{G}}=2.37$
GeV\\\hline
$\Gamma_{\tilde{G}\rightarrow\pi\pi \eta^{\prime}}/\Gamma_{\tilde{G}\rightarrow\pi\pi \eta}$ & $0.4654$ & $0.3986$\\\hline
$\Gamma_{\tilde{G}\rightarrow KK\pi}/\Gamma_{\tilde{G}\rightarrow\pi\pi \eta}$ & $0.9126$ &
$0.8553$\\\hline
$\Gamma_{\tilde{G}\rightarrow KK\eta}/\Gamma_{\tilde{G}\rightarrow\pi\pi \eta}$ & $0.0038$ &
$0.0031$\\\hline
$\Gamma_{\tilde{G}\rightarrow KK\eta^{\prime}}/\Gamma_{\tilde{G}\rightarrow\pi\pi \eta}$ &
$0.13799$ & $0.07157$\\\hline
$\Gamma_{\tilde{G}\rightarrow \eta\eta\eta}/\Gamma_{\tilde{G}\rightarrow\pi\pi \eta}$ & $0.00012$ & $0.000087$\\\hline
$\Gamma_{\tilde{G}\rightarrow \eta\eta\eta^{\prime}}/\Gamma_{\tilde{G}\rightarrow\pi\pi \eta}$ & $0.0253$ & $0.0102$\\\hline
$\Gamma_{\tilde{G}\rightarrow \eta\eta^{\prime}\eta^{\prime}}/\Gamma_{\tilde{G}\rightarrow\pi\pi \eta}$ & $0.0000012$ & $0$\\\hline
\end{tabular}%
\caption{Branching ratios for the decay of the pseudoscalar glueball $\tilde
{G}$ into three pseudoscalar mesons.}%
\end{table}%
\end{center}

\begin{center}%
\begin{table}[H] \centering
\begin{tabular}
[c]{|c|c|c|}\hline
Quantity & Case (i): $M_{\tilde{G}}=2.6$ GeV & Case (ii): $M_{\tilde{G}}=2.37$
GeV\\\hline
$\Gamma_{\tilde{G}\rightarrow\pi\pi f_{1N}}/\Gamma_{\tilde{G}\rightarrow\pi\pi \eta}$ & $0.00688$ & $0.00464$\\\hline
$\Gamma_{\tilde{G}\rightarrow KK_1\pi}/\Gamma_{\tilde{G}\rightarrow\pi\pi \eta}$ & $0.0051$ &
$0.0022$\\\hline
$\Gamma_{\tilde{G}\rightarrow K^*K^*_0\pi}/\Gamma_{\tilde{G}\rightarrow\pi\pi \eta}$ & $0.00007$ &
$0$\\\hline
$\Gamma_{\tilde{G}\rightarrow a_0\rho\pi}/\Gamma_{\tilde{G}\rightarrow\pi\pi \eta}$ & $0.0012$ &
$0$\\\hline
$\Gamma_{\tilde{G}\rightarrow a_1\eta\pi}/\Gamma_{\tilde{G}\rightarrow\pi\pi \eta}$ & $0.00289$ &
$0.00124$\\\hline
$\Gamma_{\tilde{G}\rightarrow a_1\eta^{\prime}\pi}/\Gamma_{\tilde{G}\rightarrow\pi\pi \eta}$ & $0.00019$ & $0.000001$\\\hline
$\Gamma_{\tilde{G}\rightarrow KK a_1}/\Gamma_{\tilde{G}\rightarrow\pi\pi \eta}$ &
$0.00061$ & $0.000059$\\\hline
$\Gamma_{\tilde{G}\rightarrow KK f_{1N}}/\Gamma_{\tilde{G}\rightarrow\pi\pi \eta}$ & $0.00012$ & $0.000005$\\\hline
$\Gamma_{\tilde{G}\rightarrow KK f_{1S}}/\Gamma_{\tilde{G}\rightarrow\pi\pi \eta}$ & $0.000035$ &
$0$\\\hline
$\Gamma_{\tilde{G}\rightarrow KK_1\eta}/\Gamma_{\tilde{G}\rightarrow\pi\pi \eta}$ &
$0.000009$ & $0.0000001$\\\hline
$\Gamma_{\tilde{G}\rightarrow \eta\eta f_{1N}}/\Gamma_{\tilde{G}\rightarrow\pi\pi \eta}$ & $0.000017$ &$0$\\\hline
\end{tabular}%
\caption{Branching ratios for the decay of the pseudoscalar glueball $\tilde
{G}$ into a scalar, a pseudoscalar, a vector and an axial-vector meson.}%
\end{table}%
\end{center}

In Table III, the results of the branching ratios of the pseudoscalar glueball $\tilde{G}$ for three-body decays $PPP$ are presented. Note that, these decay channels are the same decay channels that are predicted in our previous work Ref. \cite{EshraimG}, which kind of proves the validity of these decay channels.
In Table IV, we turn out to list new processes for the branching ratios of the three-body decays of pseudoscalar glueball $\tilde{G}$ into vector and axial vector mesons additional to scalar and pseudoscalar for
both choices of $M_{\tilde{G}}=2.6$ GeV and  $M_{\tilde{G}}=2.37$ GeV. That is important to widen our concept for the nature of the pseudoscalar glueball and also help researchers to detect the listed decay channels in experiments. The three-body decay channels $\pi\pi\eta$, $KK\pi$ and $\pi\pi\eta^{\prime}$ are sizable, which were also sizable in Ref.\cite{EshraimG}. Comparing Table II and III, one sees constancies such as $BR(K K \pi) > BR(K^* K)$ where the $K^* K$ is obviously a part of the $K K \pi$ channel. The decay channel $\Gamma_{\tilde{G}\rightarrow\pi\pi \pi}$ is suppressed.
The results depend only slightly on the glueball mass, which explains the similarity of their two columns.
 (For details of
the three-body decay calculation, see Sec.\ \ref{app3} of the Appendix.)

Fig. I shows the total decay action line of the pseudoscalar glueball, $\Gamma_{\tilde{G}}^{tot}=\Gamma_{\tilde{G}\rightarrow PS}+\Gamma_{\tilde{G}\rightarrow PV}+\Gamma_{\tilde{G}\rightarrow PPP}+\Gamma_{\tilde{G}\rightarrow PPA}+\Gamma_{\tilde{G}\rightarrow PPV}+\Gamma_{\tilde{G}\rightarrow PSV}$, as function of the coupling constant $c$ for both masses suggested in the present work, where the decay into baryons is negligible. The coupling constant $c$ has small value because of asymptotic freedom, and QCD decays are approximately those of free quarks and gluons at high energies.
\begin{figure}
[H]
\begin{center}
\includegraphics[
height=2.1958in,
width=3.1099in
]%
{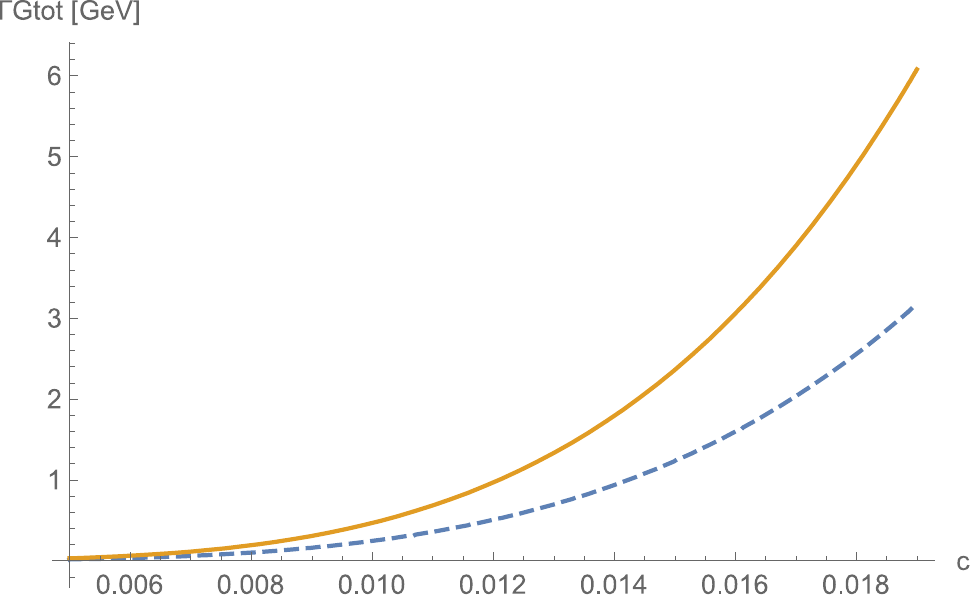}%
\caption{{\small Solid (orange) line: Total decay width of the pseudoscalar
glueball with the bare mass }${\small M}_{\tilde{G}}{\small =2.6}${\small
\ GeV as function of the coupling }$c${\small  . Dashed (blue)
line: the same curve for }${\small M}_{\tilde{G}}{\small =2.37}$ {\small GeV.}}%
\label{fig1}%
\end{center}
\end{figure}

\section{Conclusion}

The two- and three-body decays of the ground state of the pseudoscalar glueball into a vector, an axial-vector, a scalar and a pseudoscalar quark-antiquark fields have been studied. We have started with the chiral invariant effective Lagrangian describing the interaction of the pseudoscalar glueball with (axial-)vector and (pseudo)scalar mesons for the three-flavour case $N_f=3$. The size of the coupling constant intensity can not be determined. That leads to predict the branching ratios for the decay channels which are expected to dominate.  According to the mass of the pseudoscalar glueball, we considered two options: (i) $M_{\tilde{G}}=2.6$ GeV which is chosen to be in agreement with lattice QCD in the quenched approximation. (ii) $M_{\tilde{G}}=2.37$ GeV which assumes that the measured resonance $X(2370)$ in the BESIII experiment is a pseudoscalar glueball (predominantly) \cite{bes}. These masses can be tested in the planned PANDA experiment at FAIR/GSI \cite{panda}, since glueballs could be directly formed in proton-antiproton fusion processes. The two-body decay of the pseudoscalar glueball produce $PS$ (includes the scalar-isoscalar states $f_0(1370)$, $f_0(1500)$ and $f_0(1700)$) and $PV$ while the three-body decay produces $PPP,\, PPA,\, PPV$ and $PSV$. The only new two-body decay channel is $\tilde{G}  \rightarrow K^*(892)K$, see Table II for all results. The $\tilde{G} \rightarrow \pi\pi\eta$ channel is predicted in the dominant decay channel, followed by $\tilde{G} \rightarrow KK\pi$ then $\tilde{G} \rightarrow KK\eta^{\prime}$. On the contrary, the decay of the pseudoscalar glueball into $\pi\pi\pi$ is predicted to vanish. From the investigation of the decay of the pseudoscalar glueball in the present and our previous work in Ref.\cite{EshraimG}, we confirm the validity of the two- and three-body decay processes, $\tilde{G}\rightarrow PS$ and $\tilde{G}\rightarrow PPP$, that are presented in Table II and Table III, respectively. This indicates that experimentalists should search for glueballs through these channels. Moreover, the new three-body decay of the pseudoscalar glueball into (axial-)vector and (pseudo)scalar mesons is reported in Table IV. The present results of this work could be helpful for understand glueball spectroscopy and be used as guidelines in the search for the pseudoscalar glueball in the ongoing BESIII and Belle-II experiments and the future PANDA experiment at the FAIR/GSI.\\
The discovery and study of the glueball spectroscopy is a crucial test of QCD. Therefore, we plan, in the framework of a chiral model, to study the decay of the first excited pseudoscalar glueball into vector and axial-vector mesons in addition to scalar and pseudoscalar mesons, and into nucleons as well. Forthcoming developments of the present work will be based on new results for the pseudoscalar glueball and its excitations when lattice QCD will include the effect of dynamic fermions and working beyond the quenched approximation. That would be very beneficial to understand quark and gluon confinement in QCD.

\appendix

\section{The full mesonic Lagrangian}

\label{app1}

The $U(N_{f})_{L}\times U(N_{f})_{R}$ linear sigma mode \cite{dick} with (axial-)vector and (pseudo)scalar quarkonia, 
a scalar glueball $G$ and a pseudoscalar glueball $\tilde{G}$ is given by 
\begin{align}
\mathcal{L}  &
=\mathcal{L}_{dil}+\mathrm{Tr}[(D^{\mu}\Phi)^{\dagger}(D^{\mu
}\Phi)]-m_{0}^{2}\left(  \frac{G}{G_{0}}\right)  ^{2}\mathrm{Tr}(\Phi^{\dagger}%
\Phi)-\lambda_{1}[\mathrm{Tr}(\Phi^{\dagger}\Phi)]^{2}-\lambda_{2}\mathrm{Tr}(\Phi^{\dagger}%
\Phi)^{2}\nonumber\\
& +\mathrm{Tr}\left\{  \left[  \left(  \frac{G}{G_{0}}\right)
^{2}\frac {m_{1}^{2}}{2}+\Delta\right]  \left[
(L^{\mu})^{2}+(R^{\mu})^{2}\right] \right\}
-\frac{1}{4}\mathrm{Tr}[(L^{\mu\nu})^{2}+(R^{\mu\nu})^{2}]
-2\,\mathrm{Tr}[\varepsilon
\Phi^{\dagger}\Phi]\nonumber\\
&+\mathrm{Tr}[H(\Phi+\Phi^{\dagger})]+c(det\Phi-det\Phi^{\dagger})^{2}+i\tilde{c}\,\tilde{G}\left(
\text{\textrm{det}}\Phi-\text{\textrm{det}}\Phi^{\dag}\right)+\frac{h_{1}}{2}\mathrm{Tr}(\Phi^{\dagger}\Phi
)Tr[(L^{\mu})^{2}+(R^{\mu})^{2}]\nonumber\\
&+h_{2}\mathrm{Tr}[(\Phi R^{\mu})^{2}+(L^{\mu}\Phi)^{2}] +2h_{3}%
\mathrm{Tr}(\Phi R_{\mu}\Phi^{\dagger}L^{\mu})
+i\frac{g_{2}}{2}\{\mathrm{Tr}(L_{\mu\nu}[L^{\mu},L^{\nu}])+\mathrm{Tr}(R_{\mu\nu
}[R^{\mu},R^{\nu}])\}+...\text{ ,} \label{Lagc}%
\end{align}
where 
$D^\mu\Phi\equiv\partial^\mu\Phi-ig_1 (L^\mu \Phi-\Phi R^\mu)$ is the covariant derivative; and
$L^{\mu\nu}\equiv \partial^\mu L^\nu -\partial^\nu L^\mu$, and $R^{\mu\nu}\equiv\partial^\mu R^\nu -\partial^\nu R^\mu$ 
refers to the left-handed and right-handed field strength tensors. The dilaton Lagrangian

\begin{equation}
\mathcal{L}_{dil}=\frac{1}{2}(\partial_{\mu}G)^{2}-\frac{1}{4}\frac{m_{G}^{2}%
}{\Lambda^{2}}\left(
G^{4}\,\log\frac{G}{\Lambda}-\frac{G^{4}}{4}\right)\,,
\label{dil}%
\end{equation}
describes a scalar glueball $G\equiv|gg\rangle$ with quantum number $J^{PC}=0^{++}$ and mimics the trace anomaly of QCD \cite{dick, Rosenzweig, Migdal, Gomm}. The constant $\Lambda$ is the minimum of the dilaton potential which breaks the dilatation symmetry explicitly. The term $\Delta \left[(L^{\mu})^{2}+(R^{\mu})^{2}\right]$ with $\delta= \rm{diag}\{\delta_N, \,\delta_N,\,\delta_S\}$, where $\delta_i\sim m_i$ is the direct contribution of the current quark-masses to the masses of (axial-)vector mesons. It is possible to set $\delta_N=\delta_S=0$ in the isospin-symmetry limit because the identity matrix can be absorbed in the term proportional to $m_1^2$. The term $\mathrm{Tr}[H(\Phi+\Phi^{\dagger})]$ with $H=\frac{1}{2} \rm{diag}\{h_{0N},\, h_{0N},\,h_{0S}\}$, which $h_i\propto m_i^2$ is proportional to the current quark masses. Both terms break chiral symmerty due to nonzero quark masses. All the  presented paramters in the model Eq. (\ref{Lagc}) have been determined in Ref. \cite{dick}.

The following bilinear mixing terms involving the mesons
$\eta_{N}%
$-$f_{1N}$, $\overrightarrow{\pi}$-$\overrightarrow{a}_{1}$, $\eta_{S}%
$-$f_{1S}$, $K_{S}$-$K^{\ast}$, and $K$-$K_{1}$ arise \cite{dick, three
flavor}:
\begin{align}\label{mixing}
&  -g_{1}\phi_{N}(f_{1N}^{\mu}\partial_{\mu}\eta_{N}+\overrightarrow{a}%
_{1}^{\mu}\cdot\partial_{\mu}\overrightarrow{\pi})-\sqrt{2}\,g_{1}\phi
_{S}f_{1S}^{\mu}\partial_{\mu}\eta_{S}\nonumber\\&+ig_{1}(\sqrt{2}\phi_{S}-\phi
_{N})(\overline{K}\,^{\ast\mu0}\,\partial_{\mu}K_{S}^{0}+K^{\ast\mu
-}\,\partial_{\mu}K_{S}^{+})/2+\nonumber\\
&
ig_{1}(\phi_{N}-\sqrt{2}\phi_{s})(K^{\ast\mu0}\,\partial_{\mu}\overline
{K}_{S}^{0}+K^{\ast\mu+}\,\partial_{\mu}K_{S}^{-})/2\nonumber\\
&-g_{1}(\phi_{N}+\sqrt
{2}\,\phi_{S})(K_{1}^{\mu0}\,\partial_{\mu}\overline{K}^{0}+K_{1}^{\mu
+}\,\partial_{\mu}K^{-})/2\nonumber\\
&
-g_{1}(\phi_{N}+\sqrt{2}\,\phi_{S})(\overline{K}_{1}^{\mu0}\,\partial_{\mu
}K^{0}+K_{1}^{\mu-}\,\partial_{\mu}K^{+})/2\text{ .}%
\end{align}
One has to perform the following field transformations to remove the mixing terms (\ref{mixing})

\begin{equation}
\label{13}f_{1N,S}^{\mu}\rightarrow
f_{1N,S}^{\mu}+w_{f_{1N,S}}\,Z_{\eta
_{N,S}}\,\partial^{\mu}\eta_{N,S},\,\,\,\,\,\overrightarrow{a}_{1}^{\mu
}\rightarrow\overrightarrow{a}
_{1}^{\mu}+w_{a_{1}}\,Z_{\pi}\,\partial^{\mu
}\overrightarrow{\pi}\,,
\end{equation}
\begin{equation}
\label{14}K^{*\mu0}\rightarrow
K^{*\mu0}+w_{K^{*}}\,Z_{K_{S}}\,\partial^{\mu
}K^{0}_{S},\,\,\,\,\,K^{*\mu+}\rightarrow K^{*\mu+}+w_{K^{*}}\,Z_{K_{S}%
}\,\partial^{\mu}K^{+}_{S}\,,
\end{equation}
\begin{equation}
\label{15}\overline{K}\,^{*\mu0}\rightarrow\overline{K}\,^{*\mu0}+w^{*}%
_{K_{*}}\,Z_{K_{S}}\,\partial^{\mu}\overline{K}\,^{0}_{S},\,\,\,\,\,K^{*\mu
-}\rightarrow
K^{*\mu-}+w^{*}_{K_{*}}\,Z_{K_{S}}\,\partial^{\mu}K^{-}_{S}\,,
\end{equation}
\begin{equation}
\label{16}K^{\mu\pm,0}_{1}\rightarrow K^{\mu\pm,0}_{1}+w_{K_{1}}%
\,Z_{K}\,\partial^{\mu}K^{\pm,0},\,\,\,\,\,\overline{K}^{\mu0}_{1}%
\rightarrow\overline{K}^{\mu0}_{1}+w_{K_{1}}\,Z_{K}\,\partial^{\mu}%
\overline{K}^{0}\,,
\end{equation}

The constants entering into the present paper decay expressions $Z_{i}$ and $w_i$ \cite{dick} are%
\begin{equation}
Z_{\pi}=Z_{\eta_{N}}=\frac{m_{a_{1}}}{\sqrt{m_{a_{1}}^{2}-g_{1}^{2}\phi
_{N}^{2}}}\;,\,\, \,\,\,\, \, Z_{K}=\frac{2m_{K_{1}}}{\sqrt{4m_{K_{1}}^{2}-g_{1}^{2}(\phi_{N}+\sqrt{2}%
\phi_{S})^{2}}}\;,\label{zpi}%
\end{equation}%

\begin{equation}
Z_{K^*_{0}}=\frac{2m_{K^{\star}}}{\sqrt{4m_{K^{\star}}^{2}-g_{1}^{2}(\phi
_{N}-\sqrt{2}\phi_{S})^{2}}}\;,\,\, \,\,\,\, \, 
Z_{\eta_{S}}=\frac{m_{f_{1S}}}{\sqrt{m_{f_{1S}}^{2}-2g_{1}^{2}\phi_{S}^{2}}%
}\;, \label{zets}%
\end{equation}
and 
\begin{equation}
w_{f_{1N}}=w_{a_{1}}=\frac{g_{1}\phi_{N}}{m_{a_{1}}^{2}}\>,\,\,\,\,\,\,\,\,\, w_{f_{1S}}=\frac{\sqrt
{2}g_{1}\phi_{S}}{m_{f_{1S}}^{2}}\>,
\end{equation}
\begin{equation}
w_{K^{\ast}}=\frac{ig_{1}(\phi_{N}-\sqrt{2}\phi_{S})}{2\,m_{K^{\ast}}^{2}%
}\>,\,\,\, \,\,\,\,\,\,
 w_{K_{1}}=\frac{g_{1}(\phi_{N}+\sqrt{2}%
\phi_{S}}{2\,m_{K_{1}}^{2}}\,.
\end{equation}

\section{Two-body decay}

The general formula of two-body decay width is written as seen in Ref. \cite{Partig}:

\begin{equation}
\label{B1}\Gamma_{A\rightarrow BC}=\frac{S_{A\rightarrow BC}k(m_{A}%
,\,m_{B},\,m_{C})}{8 \pi m_{A}^{2}}|\mathcal{M}_{A\rightarrow
BC}|^{2},
\end{equation}
with decaying particle A and the decay products B and C. Where
\begin{equation}
k(m_{A},\,m_{B},\,m_{C})=\frac{1}{2m_{A}}\sqrt{m_{A}^{4}+(m_{B}^{2}-m_{C}%
^{2})^{2}-2m_{A}^{2}\,(m_{B}^{2}+m_{C}^{2})}\theta(m_{A}-m_{B}-m_{C}),
\label{B2}%
\end{equation}
is the center-of-mass momentum of the two
particles production in the decay,
$\mathcal{M}_{A\rightarrow BC}$ is the corresponding tree-level
decay amplitude, and $S_{A\rightarrow BC}$ refers to a
symmetrization factor (it equals $1$ if B and C are different and
it equals $1/2$ for two identical particles in the final state).\\
As an example of a two-body decay channel for $\tilde{G}\rightarrow PS$,  let us consider the case $\tilde{G}\rightarrow K K^*_0$. This process is given from Eq. (\ref{lag1}) as
\begin{equation}
\label{B1}\Gamma_{\tilde{G}\rightarrow K K^*_0}=\frac{f_{\tilde{G}\rightarrow K K^*_0} \,k_{\tilde{G} K K^*_0}}{8 \pi M_{\tilde{G}}^{2}}|-i \mathcal{M}_{\tilde{G}\rightarrow K K^*_0}|^{2},
\end{equation}

where $f_{\tilde{G}\rightarrow K K^*_0}$ is the isospin factor, $M_{\tilde{G}}$ is the pseudoscalar glueball mass, and 

\begin{equation}
|- i\mathcal{M}_{\tilde{G}\rightarrow K K^*_0}|^2= \frac{1}{4} c^2_{\tilde{G}\Phi L R }\left[\Phi_N\left(-w_{K*}\,Z_{K^*_0}+i w_{K_1}\, Z_{K}\right)
+\sqrt{2}\, \Phi_S\left(w_{K*}\,Z_{K^*_0}+\,i w_{K_1}\, Z_{K}\right) \right]^2\times\left[\frac{1}{2}\left(M_{\tilde{G}}^2-m_K^2-m_{K^*_0}^2\right)\right]^2
\end{equation}
where $m_K$ and $m_{K^*_0}$ are the masses of the kaon and $K^*_0$ mesons, respectively, while $k_{\tilde{G} K K^*_0}$ is the center of mass momentum of kaon and $K^*_0$ and reads
\begin{equation}
k_{\tilde{G} K K^*_0}=\frac{1}{2M_{\tilde{G}}}\sqrt{M_{\tilde{G}}^{4}+(m_{K}^{2}-m_{K^*_0}%
^{2})^{2}-2m_{A}^{2}\,(m_{K}^{2}+m_{K^*_0}^{2})}\,.
\label{B2}%
\end{equation}

Next we turn to the  decay of $\tilde{G}\rightarrow KK^*$ as an example for $\tilde{G}\rightarrow PV$, which its process is obtained from Eq. (\ref{lag1}) as
\begin{equation}
\Gamma_{\tilde{G}\rightarrow KK^*}=\frac{f_{\tilde{G}\rightarrow KK^*} \,k_{\tilde{G} KK^*}}{8 \pi M_{\tilde{G}}^{2}}.\frac{1}{3}\left[\frac{1}{2}c_{\tilde{G}\Phi L R }(\Phi_N-\sqrt{2}\Phi_S)\right]^{2}\left[- m_K^2+ \frac{M_{\tilde{G}}^2-m_{K^*}^2-m_K^2}{2m_{K^*}^2}\right]^2\,,
\end{equation}
where $f_{\tilde{G}\rightarrow KK^*}=4$ and $k_{\tilde{G} KK^*}$ is the center of mass momentum of $K$ and $K^*$. In an analogous way, all the decay processes $\tilde{G}\rightarrow PV$ in table I are calculated the corresponding change of the isospin factors, the decay products mass and the constants entering in the amplitudes.

\section{Three-body decay}

\label{app3}

The general explicit expression for the three-body decay width for the process $A\rightarrow B_{1}B_{2}B_{3}$ reads \cite{PDG}:
\[
\Gamma_{\tilde{G}\rightarrow P_{1}P_{2}P_{3}}=\frac{f_{\tilde{G}\rightarrow
P_{1}P_{2}P_{3}}}{32(2\pi)^{3}M_{\tilde{G}}^{3}}\int_{(m_{1}+m_{2})^{2}%
}^{(M_{\tilde{G}}-m_{3})^{2}}dm_{12}^{2}\int_{(m_{23})_{\min}}^{(m_{23}%
)_{\max}}|-i\mathcal{M}_{\tilde{G}\rightarrow P_{1}P_{2}P_{3}}|^{2}dm_{23}^{2}%
\]
where
\begin{align}
(m_{23})_{\min}  &  =(E_{2}^{\ast}+E_{3}^{\ast})^{2}-\left(  \sqrt{E_{2}%
^{\ast2}-m_{2}^{2}}+\sqrt{E_{3}^{\ast2}-m_{3}^{2}}\right)  ^{2}\text{ ,}\\
(m_{23})_{\max}  &  =(E_{2}^{\ast}+E_{3}^{\ast})^{2}-\left(  \sqrt{E_{2}%
^{\ast2}-m_{2}^{2}}-\sqrt{E_{3}^{\ast2}-m_{3}^{2}}\right)  ^{2}\text{ ,}%
\end{align}
and%
\begin{equation}
E_{2}^{\ast}=\frac{m_{12}^{2}-m_{1}^{2}+m_{2}^{2}}{2m_{12}}\text{ , }%
E_{3}^{\ast}=\frac{M_{\tilde{G}}^{2}-m_{12}^{2}-m_{3}^{2}}{2m_{12}}\text{ .}%
\end{equation}
The quantities $m_{1},$ $m_{2},$ $m_{3}$ are the masses of the three decay products $P_{1},$ $P_{2}$, and $P_{3},$ $\mathcal{M}_{\tilde
{G}\rightarrow P_{1}P_{2}P_{3}}$ is the corresponding tree-level decay
amplitude, and $f_{\tilde{G}\rightarrow P_{1}P_{2}P_{3}}$ is a symmetrization
factor which equals $1$ if all decay products are different,
equals $2$ for two identical decay products in the final state, and it equals $6$ if  $P_{1},$ $P_{2}$, and $P_{3},$
are identical in the final state.\\

For example, the amplitude for the process $\tilde{G}\rightarrow\overline K^{*0}K^{*0}_0 \pi^0$ is 

\begin{equation}
|- i\mathcal{M}_{\tilde{G}\rightarrow {K}^{*0}K^{*0}_0 \pi^0}|^2= \frac{c^2_{\tilde{G}\Phi L R }}{4} . \frac{1}{3}\left[-\left(Z_{\pi} m_{K^*_0}^2+ 2 Z_\pi Z_{K^*_0} k_\pi.k_{K^*_0}+m_\pi^2 \, Z_{K^*_0}^2\right)+\frac{1}{m^2_{K^*}}\left(Z_\pi k_{\overline{K}^{*0}}.k_{K^*_0}+Z_{K^*_0} k_{\overline{K}^{*0}}.k_{\pi^0}\right)^2\right]\,,
\end{equation}
where
\begin{align}
 k_{\overline{K}^{*0}}.k_{\pi} &  =\frac{m^2_{12}-m^2_{\overline{K}^{*0}}-m^2_{K^*_0}}{2}\text{ ,}\\
 k_\pi.k_{K^*_0}&=\frac{m^2_{23}-m^2_{\pi}-m^2_{K^*_0}}{2}\text{ ,}\\
 k_{\overline{K}^{*0}}.k_{K^*_0}  &  =\frac{m^2_{13}-m^2_{\overline{K}^{*0}}-m^2_{\pi}}{2}\,.
\end{align}

The other three-body decays are calculated in analogous way.\\

Notice that there are several decay channels of the pseudoscalar glueball, $\tilde{G}$ that appear in Eq. ($\ref{lag1}$) but they are not kinematically allowed because the mass of the decaying particle is lower than the sum of the mass of the decay products  $M<\sum^3_i\,\, m_i$.

\end{document}